\documentclass[useAMS,usenatbib]{mn2e}
\usepackage{url}
\usepackage{stmaryrd}
\usepackage[pdftex]{graphicx}
\usepackage[draft]{hyperref}

\def\aj{AJ}%
\def\apj{ApJ}%
\def\apjl{ApJ}%
\def\apjs{ApJS}%
\def\apss{Ap\&SS}%
\def\aap{A\&A}%
\def\mnras{MNRAS}%
\def\pasp{PASP}%
\def\sns{SNe\,{\sc I}a}
\def\sn{SN\,{\sc I}a}
\def\av{$A_V$}
\def\nSN{3585}

\begin{document}

\title{SN\,Ia Host Galaxy Properties and the Dust Extinction Distribution}



\author[Holwerda et al.]{
B.W. Holwerda$^{1}$\thanks{E-mail: holwerda@strw.leidenuniv.nl, twitter: @benneholwerda},
A. Reynolds$^{2}$,
M. Smith$^{2}$, and R. C. Kraan-Korteweg$^{2}$\\
$^{1}$ University of Leiden, Sterrenwacht Leiden, Niels Bohrweg 2, NL-2333 CA Leiden, The Netherlands\\
$^{2}$ Astrophysics, Cosmology and Gravity Centre ($AC/GC$), \\
Astronomy Department, University of Cape Town, Private Bag X3, 7700 Rondebosch, Republic of South Africa\\
}

\date{Accepted ----. Received ----; in original form ----}

\pagerange{\pageref{firstpage}--\pageref{lastpage}} \pubyear{2002}

\maketitle

\label{firstpage}

\begin{abstract}
Supernovae Type {\sc I}a display a complex relation with their host galaxies. An important prior to the fit of the supernovae's lightcurve is the distribution of host galaxy extinction values that can be encountered. The SDSS-SN project has published light curve fits using both MLCS2k2 and SALT2. We use the former fits extinction parameter ($A_V$) to map this distribution of extinction values.

We explore the dependence of this distribution on four observables; the inclination of the host galaxy disk, radial position of the supernova, redshift of the supernova and host, and the level of star-formation in the host galaxy. 
The distribution of \av \ values encountered by supernovae is typically characterised by :$\rm N_0 ~ e^{-A_V/\tau}$, with $\tau$= 0.4 or 0.33.

We find that the inclination correction using an infinitely thin disk for the \sn \  is sufficient, resulting in similar exponential \av\ distributions for high- and low-inclination disks. The \av \ distribution also depends on the radial position in the disk, consistent with previous results on the transparency of spiral disks. 
The distribution of \av \ values narrows with increased star-formation, possibly due to the destruction or dispersion of the dusty ISM by stellar winds prior to the ignition of the supernova.

In future supernova searches, certainly the inclination of the host galaxy disk, should be considered in the construction of the \av \ prior with $\tau=0.4/cos(i)$ as the most likely prior in each individual host galaxy's case.
\end{abstract}


\section{\label{s:intro}Introduction}

Interest in type Ia supernovae (SNe {\sc I}a) has greatly increased since their use as standard candles led to the discovery of the accelerated expansion of the Universe \citep{Riess98,Perlmutter99}. The latest generation of \sn~ surveys, e.g., the Supernova Legacy Survey \citep[SNLS,][]{Astier06}, Equation
of State: SupErNovae trace Cosmic Expansion \citep[ESSENCE,][]{Wood-Vasey07}, and the Sloan Digital Sky Survey SuperNova survey \citep[SDSS-SN,][]{Kessler09}, however, are limited by the possibility of systematic uncertainties in their calibration of \sn \  luminosities. In order for \sn \  light curves to evolve into a next generation cosmological tool, we will need to understand the various physical properties that could affect the relation between peak luminosity and light curve width \citep{Phillips93} that is used to calibrate SNe {\sc I}a. 

Many studies focus on the stellar population which produces the supernova. For example, \cite{Gallagher05, Sullivan06, Mannucci06, Pan13a, Kistler13} derive host galaxy stellar population and metallicity from general photometry to compare these to the \sn \  rates and luminosity profiles. In general, a two sub-population model is emerging: blue, star forming galaxies host higher rates of fast declining supernova and red, passive galaxies host predominantly more slowly-declining SNe~Ia \citep{Hamuy96, Howell01, van-den-Bergh05, Mannucci06, Schawinski09, Lampeitl10b, Wang13a}. 
This model suggests the difference in light curve characteristics to the different progenitor stellar populations. 

However, simultaneously, the issue of host galaxy extinction has come to the fore, either the applicability of the appropriate extinction law \citep[][]{Riess96b,Phillips99,Altavilla04,Reindl05, Jha07, Conley07}, the validity of the prior of extinction values \citep{Jha07, Wood-Vasey07}, or the possibility that the geometry of the dusty ISM in (and hence the extinction distribution) may change with galaxy evolution \citep{Holwerda08}. The source of extinction in supernovae is thought to be the surrounding host galaxy \citep{Phillips13a}.
The uncertainty in host galaxy extinction has been identified in the Dark Energy Task Group Report \citep{DETF} as the dominant remaining systematic that stands in the way of  \sn \ measurements attaining the next step in cosmological precision. Current attention is on the photometric calibration of the various surveys into a single system \citep[e.g.,][]{Conley11, Scolnic14, Betoule14} but after this technical issue is dealt with the host extinction still stands in the way of the final goal of 1\% precision for SN\,I \citep[e.g.,][]{Riess11,Kelly14}.

The distribution of host galaxy extinction values is an important Baysian prior to the MLCS2k2 lightcurve fit program \citep{Kessler09,Jha07}. 
The benefits of the use of a Baysian prior are (1) full use of the information on the host galaxy to reduce the uncertainty, and (2) mitigation of the effect of colour measurement error \citep[see the appendix in][]{Jha07}. A distinct disadvantage is that an incorrect prior will result in biases in the distance measurement and some light curve fitting programs therefore have opted to not include a host galaxy extinction prior.
To date, the extinction model is primarily based on Monte-Carlo simulations of spiral disks \citep{Hatano98,Commins04, Riello05} that also predicted \sn \  rates but check against the data \citep[][and their Figure 6]{Jha07}. 
The dusty disk in these model host galaxies was assumed to be extremely flat, relatively transparent with some approximation for the spiral arms. 
These models may now be in need of revision as observations of nearby spirals show a much more complex dust geometry of spiral galaxies: part of the dust is distributed in a vertical component with a similar height to the stellar disk \citep[e.g.,][]{Howk97,Howk99a,Howk99b,Thompson04,Seth05b,Kamphuis07}, and the radial distribution is increasingly better characterised with a disk and radially declining spiral arm component \citep[see,][]{kw99a,kw00a,kw00b,kw01a,kw01b, Keel13, Holwerda05a, Holwerda05b, Holwerda05c, Holwerda05e, Holwerda05d, Holwerda07a, Holwerda07b, Holwerda07c, Holwerda09a, Holwerda13a, Holwerda13b}. 

Luckily, there is an opportunity to use the \sn \ themselves to probe the host galaxy dust disk characteristics, as large samples of supernova in the nearby Universe, complete with host galaxy properties, are becoming available, notably the Sloan Digital Sky Survey-SN Supernova (SDSS-SN) search.
 In this paper, we compare the distribution of dust extinction values (\av) for the supernova in this sample \citep[S14 hereafter][]{Frieman08, Sako08, Sako14} for which uniform host galaxy properties (inclination, positions, typical radii, a classification of star formation and redshift are available. 

Our goal here is not to present a precise measurement of the underlying $A_V$ distribution, but rather to explore how the observed distribution relates to host galaxy observables to ascertain which host galaxy property should have priority in future SN\,Ia surveys.
The SDSS-SN fits used in this analysis made use of an particular extinction distribution as a prior, and we are arguing that this prior was at best the average of the priors of different host galaxy populations. However, as long as the different host galaxy populations have $A_V$ distribution derived using the same prior, the use of an incorrect prior should only decrease our sensitivity to detect a relationship with a host galaxy characteristic rather than introduce a spurious relation, and hence does not invalidate our results. Future, more detailed studies aiming to precisely measure the underlying $A_V$ distribution will have to properly take the effects of the S14 prior into account.

This paper is organized as follows: 
section \ref{s:data} describes the SDSS-SN data we used for this paper,
section \ref{s:analysis} describes our analysis in detail, section \ref{s:concl} lists our conclusions.

\section{Data}
\label{s:data}

Supernovae type Ia (SN\,Ia) are the mainstay of Cosmological distance measurements. The photometric properties of their light curves are very stable but appear to depend slightly on host galaxy properties \citep[e.g., star-formation rate or stellar mass][]{}.
With this mind, the third incarnation of the Sloan Digital Sky Survey (SDSS-3) included a supernova search with spectroscopic follow-up. This SDSS-Supernova search (SDSS-SN)\footnote{\url{http://www.sdss.org/supernova/aboutsupernova.html}}  has yielded a wealth of galaxy and supernova properties.

SDSS-SN is described in \cite{Frieman08} and in more detail in \cite{Sako08} and \cite{Sako14} (S14). 
Observing for three months of the year from 2005 to 2007, SDSS-SN identified hundreds
spectroscopically confirmed SNe \,Ia in the redshift range 0.05 $<$ z $<$ 0.35. 

S14 present the final data-product of the tremendous observational effort. They include an assessment of SN type based on the light curve (PSNID output) and light curve fits using the two most commonly used packages, SALT2 \citep{Guy07} and  MLCS2k2 \citep{Jha07}.
The shape of the MLCS2k2 $A_V$ prior used for the light curve fits was: $P(A_V) \sim e^{-A_V/\tau_0}$, with $\tau_0 =0.4$. Because it has host galaxy extinction as an explicit prior and a model relation between $A_V$ distribution width ($\tau_0$) and $A_V$ bias, we use the MLCS2k2 values for our further analysis here.
For example, \citep{Wood-Vasey07} use this tool to explore the dependence of the extinction law used on redshift. 
We focus only on  those objects that have a reasonable chance of being bona-fide SN\,Ia (S/N$>$5 and P(SN\,Ia) $>$ 90\%, according to PSNID) to ensure the conclusions for the improved prior are based on these only (\nSN\ SN in the total sample).

Host galaxy properties are those available from the SDSS-DR9 database and stellar mass and star-formation are modeled with two different packages, FSPS \citep{FSPS} and PEGASE \citep{Bruzual09}. For the purpose of this paper, the star formation is of interest and we choose the FSPS value for for the further analysis. 

In addition to the values thoughtfully provided by the SDSS-SN project in S14, we retrieve the axes ratio, position angle, and petrosian radius from the SDSS server. Using these values, we compute the galactocentric radial distance from the center of the host galaxy in Petrosian radii. 

\section{Analysis}
\label{s:analysis}

The analysis focuses on four observables of the host galaxy and their effect on the \av \ distribution: disk inclination ($i$), radial distance from the centre of the galaxy disk ($r$), redshift ($z$), and the level of star-formation (type); of these the last has been mostly the focus of the SDSS-SN collaboration \citep{Lampeitl10b}. 
The SN sample has a complex selection function, part of which is the extinction \av \ that we are interested in. Hence we compare subsamples of the SDSS-SN sample; in this case the selection is at least uniform, if not always complete.
We compare subsamples of SNe {\sc I}a based on these four host galaxy characteristics and plot the distribution of \av \ values, typically with 25 bins. We fit an exponential to each \av \ distribution to quantify the drop-off rate:
\begin{equation}
N = N_0 e^{-A_V/\tau},
\label{eq:exp}
\end{equation}
\noindent for positive values of $A_V$. 
The typical \sn \  prior value for $\tau$ is 0.33 \citep{Kessler09}, 0.4 \citep{Jha07,Sako08} or $0.38\pm0.06$ \citep{Bernstein11}. The mean extinction law is found to be $R_V = 2.3$, which is lower than the canonical Milky Way extinction law \citep[$R_V =3.1$,][]{ccm}.
In their appendix, \cite{Lampeitl10b} try to constrain this $R_V$ value using this SN sample. 

The central assumptions for our analysis is that 
(a) the shape of the $A_V$ distribution is similar to the prior shape (equation \ref{eq:exp}), i.e., the convolution with data does not result in a dramatic change,
and 
(b) the $A_V$ distribution for subsamples selected based on host characteristics that do not matter to the $A_V$ distribution should be the same.

In the following sections, we fit an exponentially modified Gaussian distribution to the full \av\ distributions using a maximum likelihood method, normalizing it arbitrarily for visualization purposes. In each panel, we list the exponential drop-off, its uncertainty and the number of $A_V$ values in the distribution.

Our second assumption is that above \av=1, the sample is not complete enough to contribute to a meaningful fit. The optimal number of bins (15) in the histograms was chosen by eye. 

We should point out that both this paper and all the \sn\ work has assumed so far that the probability function for $A_V$ follows equation \ref{eq:exp}. However, spectral energy distribution fits of stars in M31 (Dalcanton et al. {\em in preparation}), column density distributions of star-forming regions and our own HST mapping of extinction values all hint that a log-normal distribution may well be a better description of the the data. 

\subsection{Inclination}
\label{s:incl}

Inclination of a spiral disk can be found from the major (a) and minor (b) axis according to:
\begin{equation}
\label{eq:i}
\cos^2(i) = { (b/a)^2  - (b/a)^2_{min}\over (b/a)^2_{min} }
\end{equation}
\noindent from \cite{Hubble26}. The value for $(b/a)_{min}$ is the typical  oblateness of an edge-on galaxy ($b/a_{min} = 0.1$). The axes are those observed for the Petrosian aperture in the sdss-{\em r} filter. Hence the oblateness of the dusty ISM disk may well be different from the $(b/a)_{min}$ used here. Typically, it is assumed that this dusty ISM disk is in fact much thinner than the stellar disk ($(b/a)_{min}\sim0.1$). We apply this correction regardless whether the galaxy is an early or late-type because (a) the majority of host galaxies is late-type, and (b) the ISM will still be in a disk in the case of early-types.

\begin{figure}
\begin{center}
\includegraphics[width=0.5\textwidth]{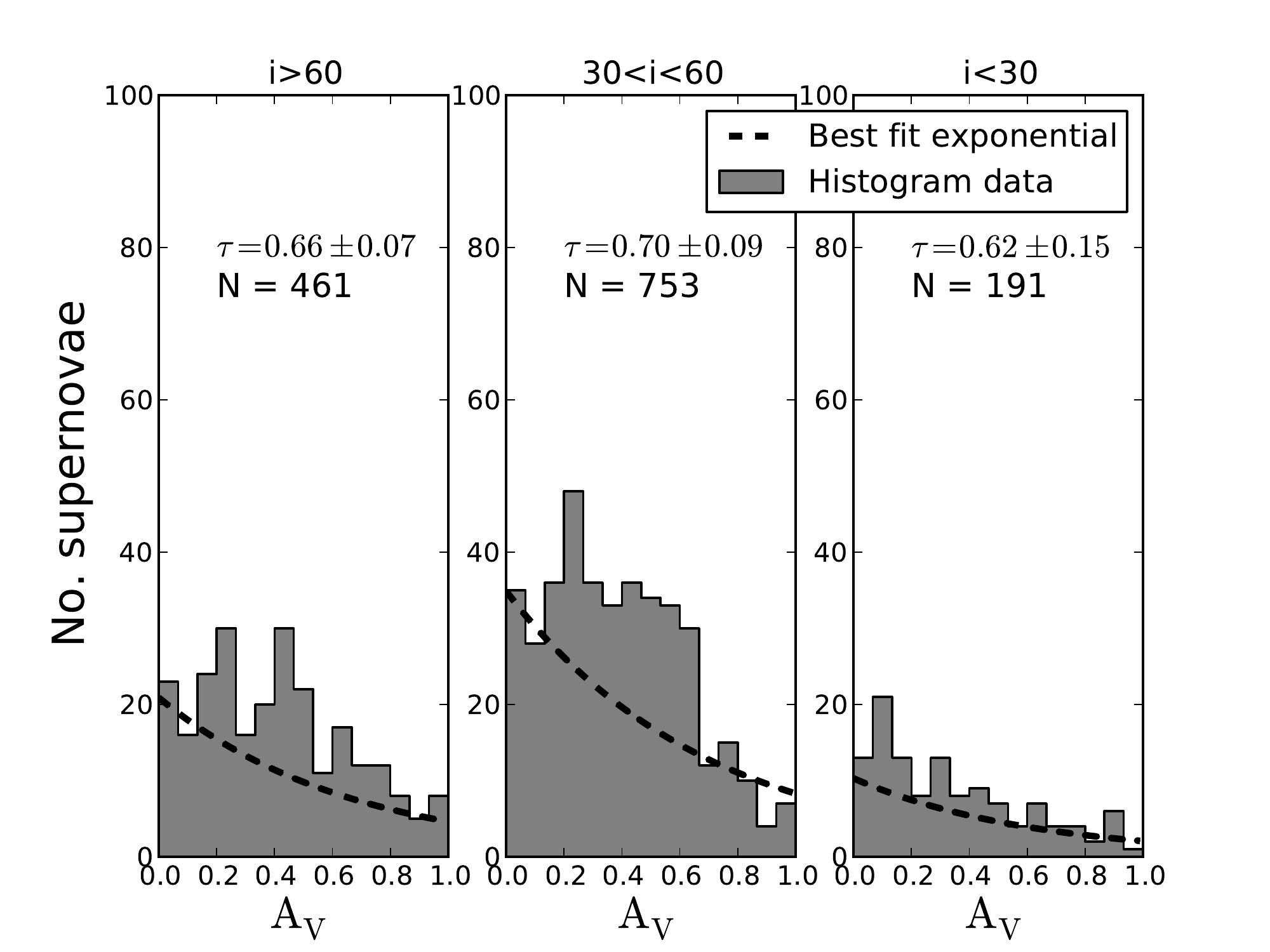}
\caption{The distribution of observed $A_V$ values for the high- ($i>60^\circ$), mid- ($30^\circ < i < 60^\circ$), and low-inclination ($i<30^\circ$) galaxies, before correction for inclination. In each panel, we list the exponential drop-off, its uncertainty and the number of $A_V$ values in the distribution.}
\label{f:obs}
\end{center}
\end{figure}

\begin{figure}
\begin{center}
\includegraphics[width=0.5\textwidth]{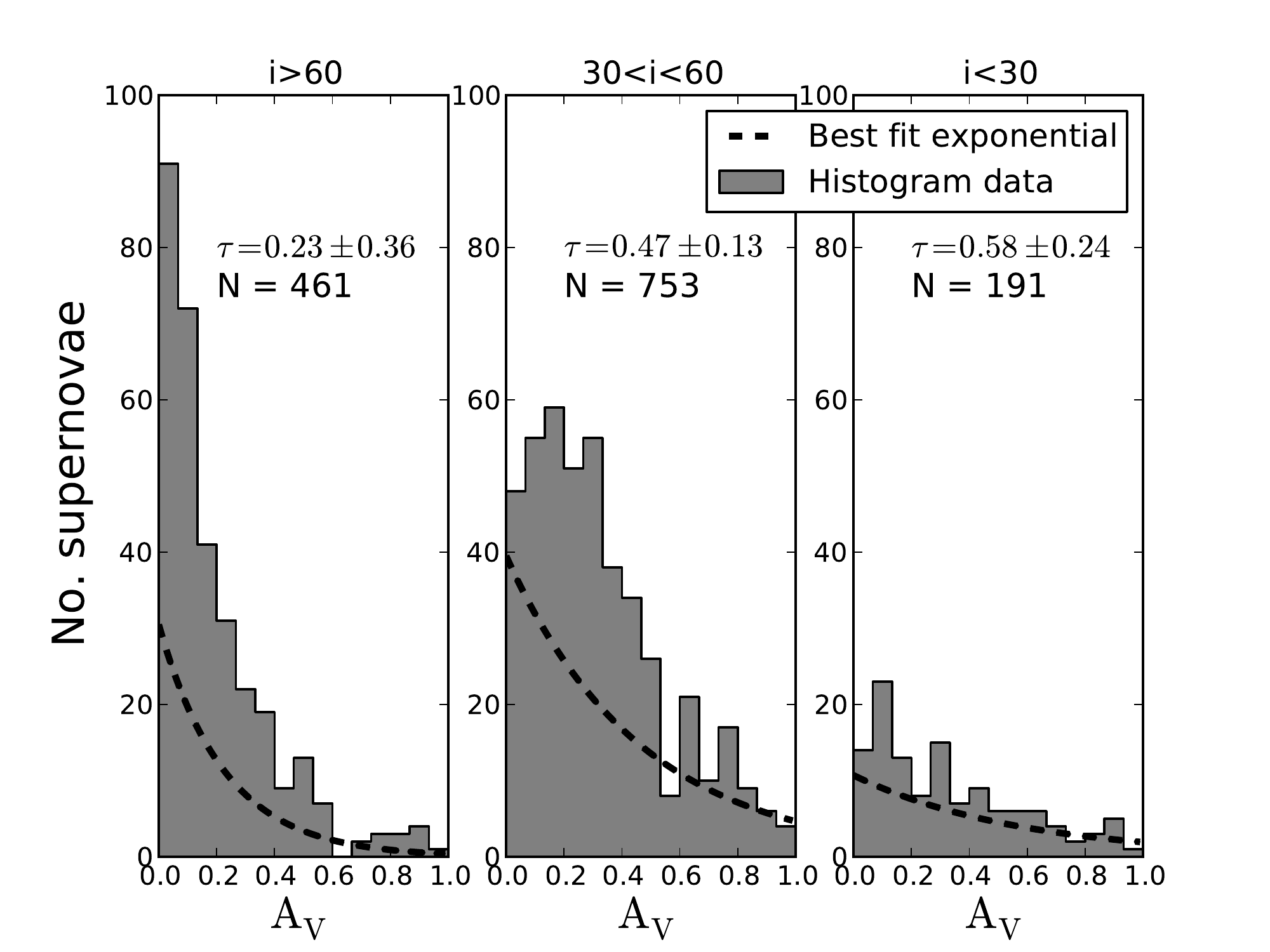}
\caption{The distribution of corrected $A_V$ values for the high- ($i>60^\circ$), mid- ($30^\circ < i < 60^\circ$), and low-inclination ($i<30^\circ$) galaxies. Drop-off values are more similar now for the moderately and face-on inclined.}
\label{f:cor}
\end{center}
\end{figure}

\begin{figure*}
\begin{center}
\includegraphics[width=0.49\textwidth]{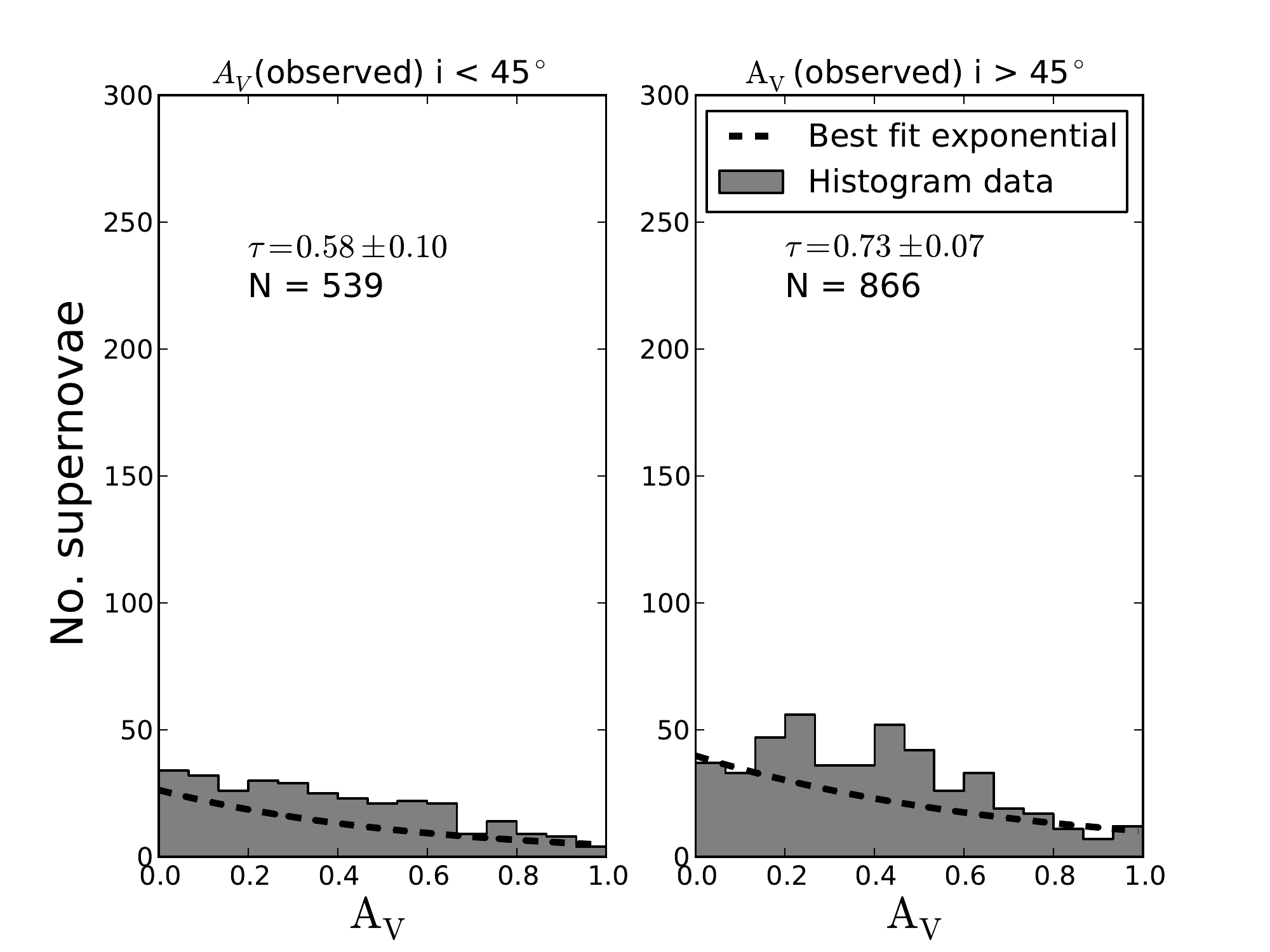}
\includegraphics[width=0.49\textwidth]{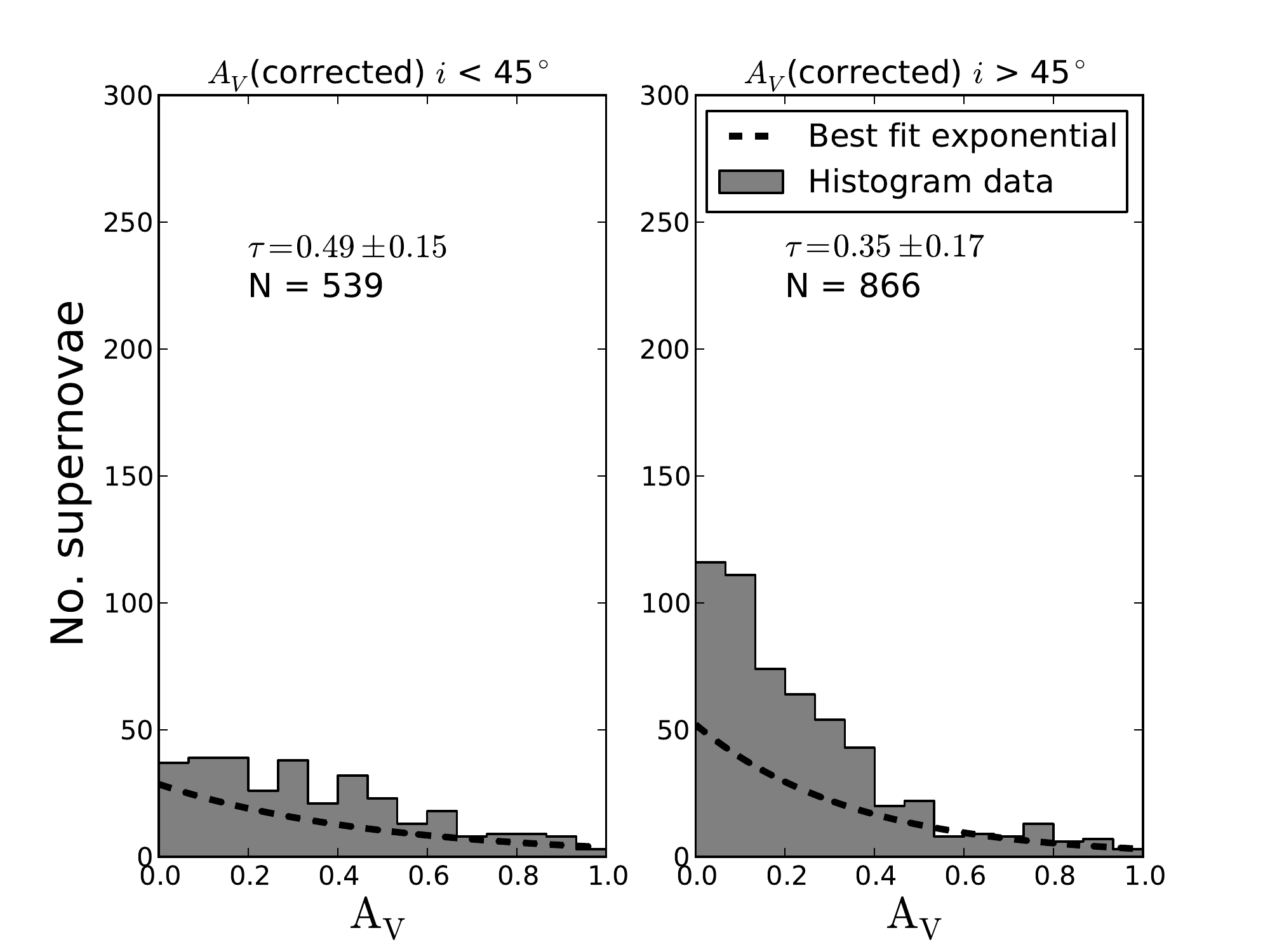}
\caption{The distribution of corrected $A_V$ values for the high- ($i>60^\circ$), mid- ($30^\circ < i < 60^\circ$), and low-inclination ($i<30^\circ$) galaxies. Drop-off values are more similar now for the moderately and face-on inclined.}
\label{f:correction}
\end{center}
\end{figure*}

\begin{figure}
\begin{center}
\includegraphics[width=0.5\textwidth]{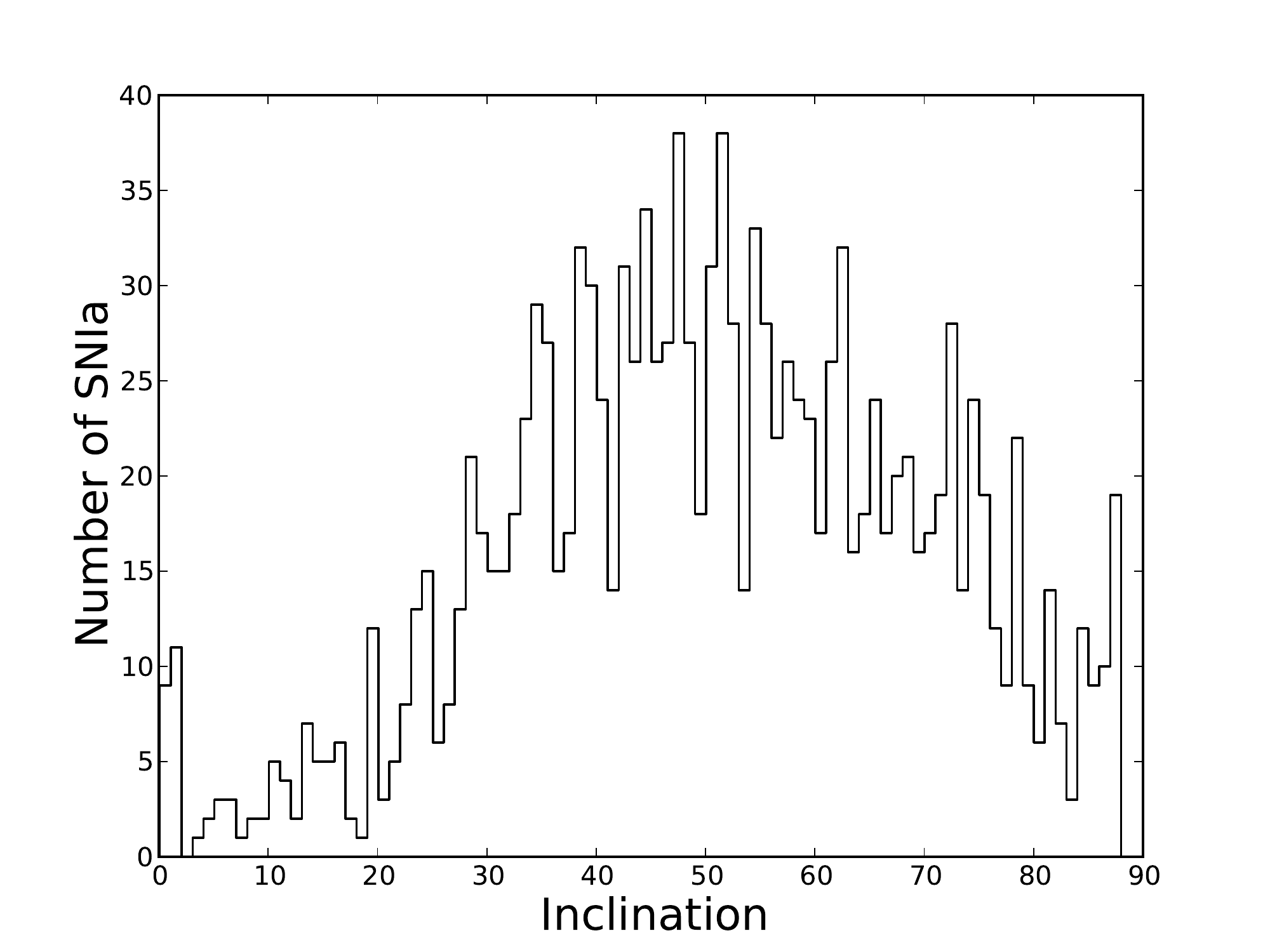}
\caption{The distribution of disk inclination of the host galaxies based on the axis ratio ($B/A$).}
\label{f:hist:incl}
\end{center}
\end{figure}

In Figures \ref{f:obs}, \ref{f:cor} and \ref{f:correction}, we compare the \av \ distribution of high- and low-inclination disks, before and after correction of the \av \ values for inclination, assuming a thin disk model (b/a$_{min}$ = 0.1). We note that the majority of our host galaxies are moderately inclined ($30^\circ<i<60^\circ$, Figure \ref{f:hist:incl}), as can be expected from a random draw. The correction using this simple relation with inclination (cos(i) = b/a) is sufficient to align the two distributions (similar $\tau$ values). Hence, nothing fancier is needed. We did vary the $(b/a)_{min}$ value, but did not obtain a significantly better result. 

An extremely thin dusty ISM disk is all that is needed to model the dusty ISM distribution. This is somewhat unexpected at intermediate redshift because of the evidence for a secondary vertical dusty ISM component \citep[e.g.,][]{Kamphuis07,Bianchi11,Holwerda12a, Holwerda12b, Schechtman-Rook13,Seon14}. It may be even less applicable at higher redshift; for instance, with an order of magnitude more star-formation at $z\sim1$, the resulting turbulence could result in a much thicker distribution at higher redshifts \citep{Holwerda08}. For the remainder of this letter, we use the inclination corrected \av \ values. 

\begin{figure}
\begin{center}
\includegraphics[width=0.5\textwidth]{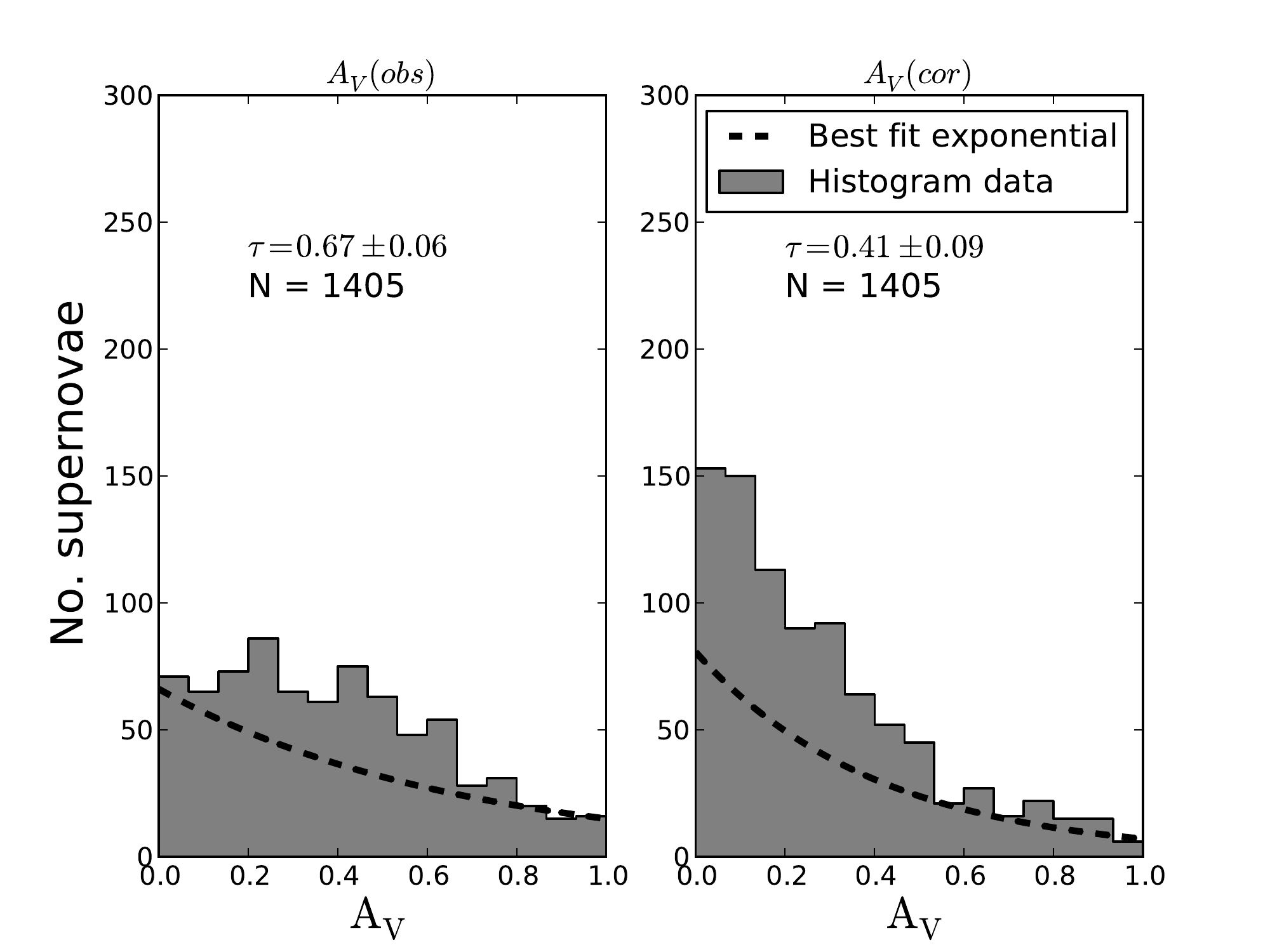}
\caption{The observed $A_V$ distribution and the $A_V$ distribution after correction for disk inclination. }
\label{f:AVcor}
\end{center}
\end{figure}

Figures \ref{f:correction} and \ref{f:AVcor} shows the distribution of $A_V$ values before and after correction for inclination. While the {\sc MLCS2k2} package can start with a {\em flat} prior for the dust attenuation, the distribution is wider than typically assumed for the prior, however, after correction to face-on, it is practically identical to the values ($\tau=0.33,0.4$) typically in use. We advocate therefore for the use of an extinction prior tailored using the known host galaxy inclination: $\tau\sim0.4/cos(i)$, where $i$ is the disk's inclination.

\subsection{Radius}

\begin{figure}
    \centering
    \includegraphics[width=0.5\textwidth]{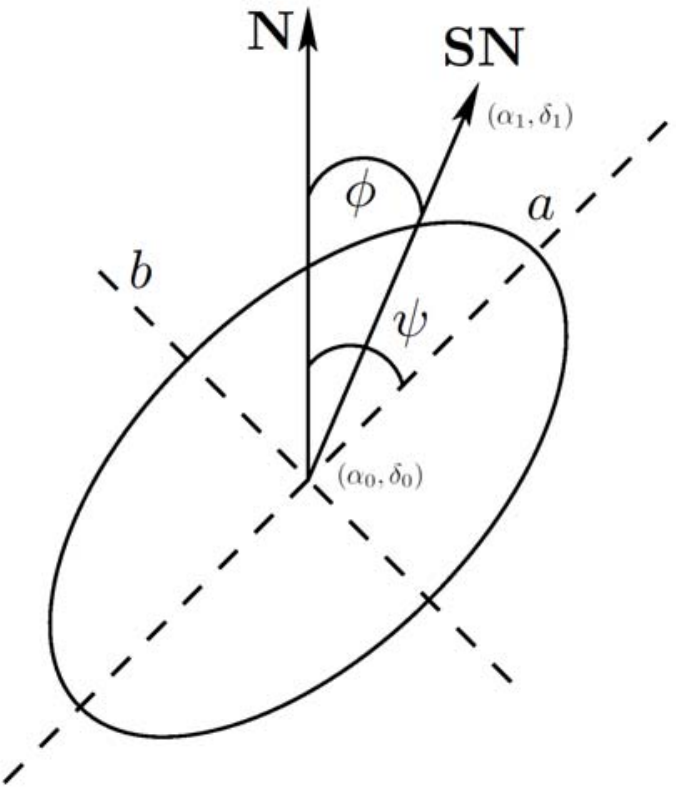}
    \caption{Geometry of the host galaxy and SN on the plane of the sky. The centre of the host galaxy ($\alpha_0,\delta_0$), and the supernova ($\alpha_1,\delta_1$),}
    \label{f:projection}
\end{figure}

Secondly, we test the dependence of the \av \ distribution on the radial distance from the centre of the host galaxy. The radial distance can be obtained from the position of the galaxy ($\alpha_0,\delta_0$), the position of the supernova ($\alpha_1,\delta_1$), the inclination ($i$), the position angle on the sky ($\psi$), and the angle between the two positions ($\phi$):
Figure \ref{f:projection} shows the the projected position of the supernova on the inclined disk of the spiral host galaxy. 

Using the expressions for proper motion, one can find the de-projected radial distance from the centre of the galaxy:
\begin{equation}
R = \mu\sqrt{\left[ \frac{\sin{(\phi - \psi)}}{\cos{i}}\right]^2 + \cos^2{(\phi - \psi)}}
\end{equation}
which, combined with the Petrosian radius \citep{Petrosian76}, can be used to divide the \sn \  up into samples inside and outside the stellar disk of the host galaxy. We use the Petrosian radius to scale the radial positions of the SNe {\sc I}a in all the galaxies. 

\begin{figure}
\begin{center}
\includegraphics[width=0.5\textwidth]{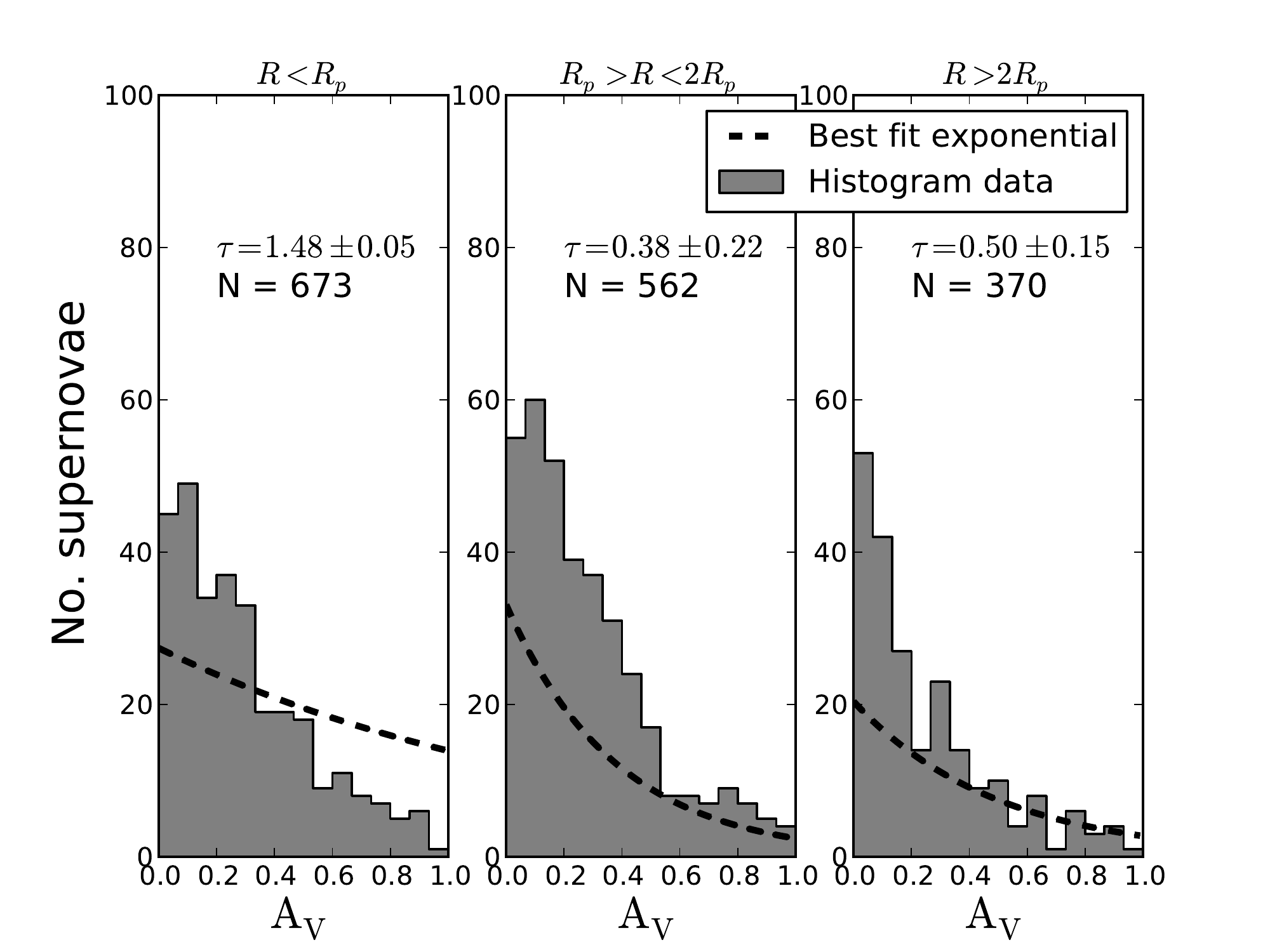}
\caption{The distribution of inclination corrected $A_V$ values for three devisions according to radius, expressed as a fraction of the Petrosian Radius of the host galaxies.}
\label{f:Rp}
\end{center}
\end{figure}

\begin{figure}
\begin{center}
\includegraphics[width=0.5\textwidth]{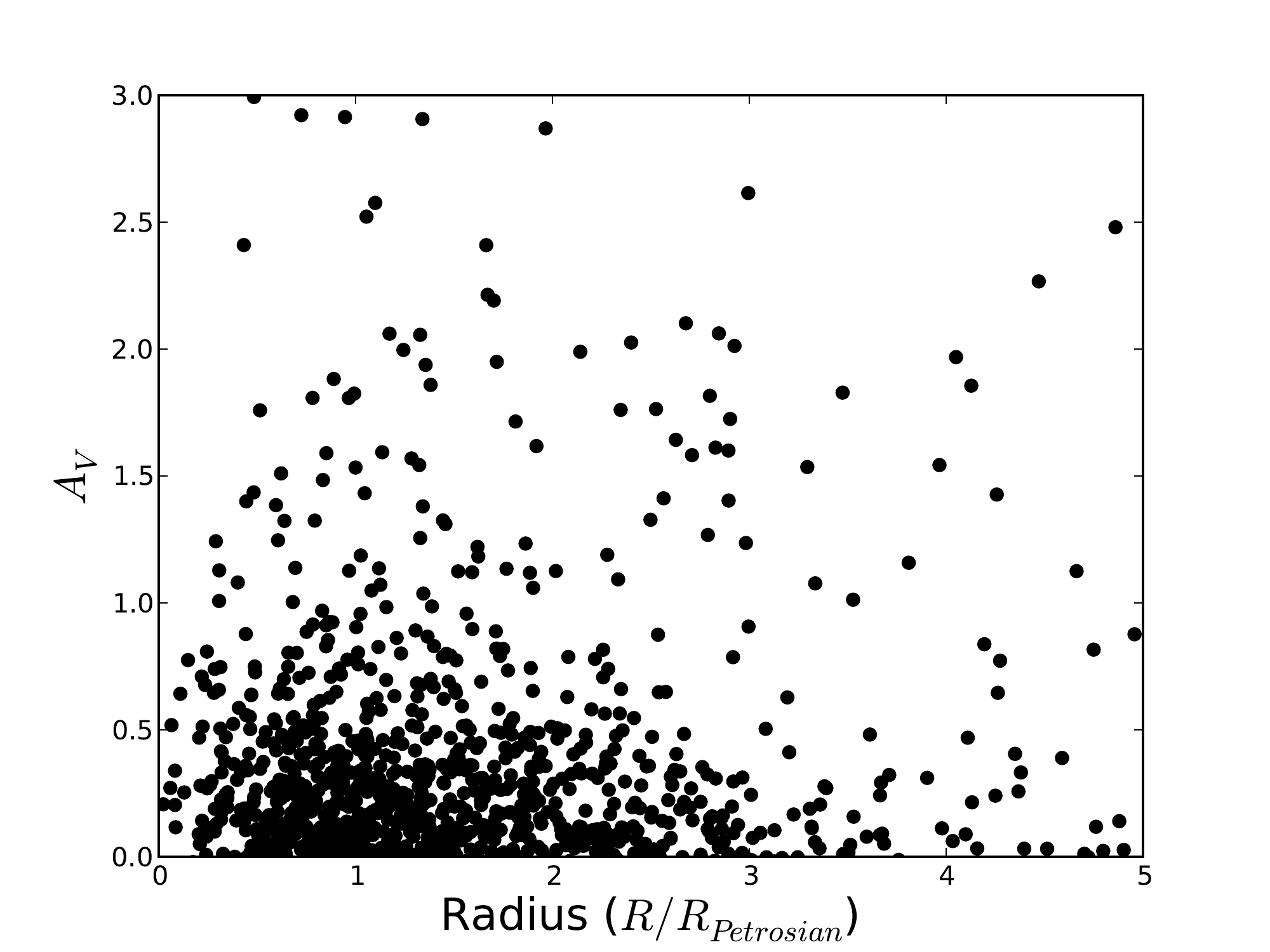}
\caption{The inclination corrected $A_V$ values, as a function of radius of the host galaxies, expressed as a fraction of the Petrosian radius. Three of the \sns \  are beyond five Petrosian radii and could have due to misidentified host galaxies. These were excluded for further analysis.}
\label{f:RAv}
\end{center}
\end{figure}

Figure \ref{f:Rp} shows three \av \ distributions within one, between one and two times the Petrosian radius and outside two Petrosian radii. Supernovae inside two Petrosian radii encounter a variety of extinction values ($\tau \sim 1.48--0.38$), as can be expected and those outside two Petrosian radii encounter a more extended distribution ($\tau \sim 0.5$). Hence, where the \sn \  is in the plane of the galaxy is critical for determining the \av \ distribution to be used. Figure \ref{f:RAv} shows the radial distribution of \av \ values, normalized to the Petrosian radius and it shows a consistent picture of an extended disk of \av=0.5 out to greater radii and a steeper falling off spiral arm component \citep[consistent with the disk transparency measures presented in][]{kw99a, kw00b,kw01a, kw01b, Keel13, Keel14, Holwerda05b, Holwerda07c, Holwerda09a, Holwerda13a, Holwerda13b}.

\subsection{Redshift}

\begin{figure}
\begin{center}
\includegraphics[width=0.5\textwidth]{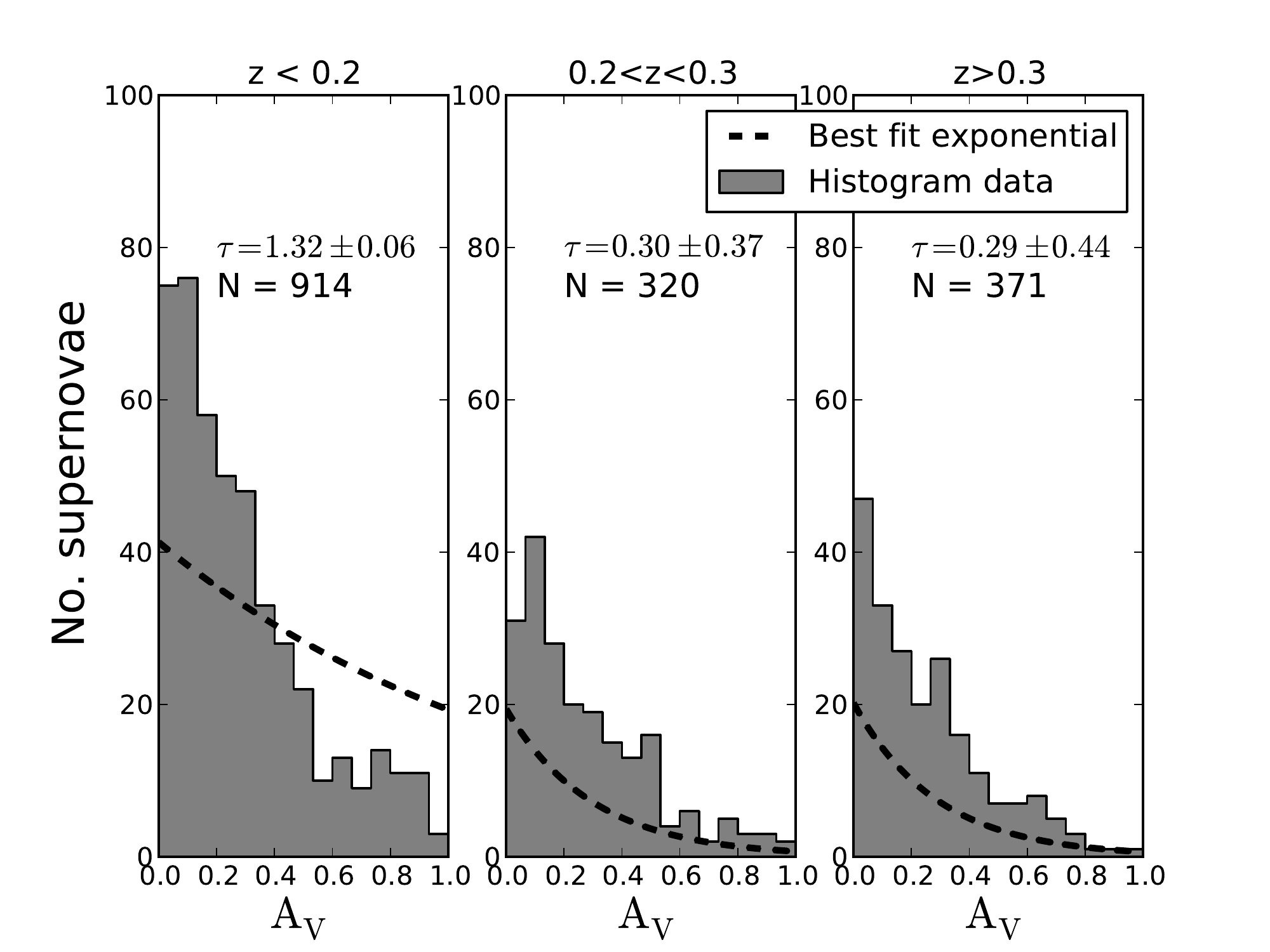}
\caption{The distribution of inclination corrected $A_V$ values for different redhift ranges.}
\label{f:z}
\end{center}
\end{figure}

The third characteristic of the host galaxy that may influence the \av \ distribution is evolution of the host galaxy's dust geometry \citep[see][]{Holwerda08}. The selection effects as a function of redshift was exhaustively explored in \cite{Wood-Vasey07} and \cite{Kessler09}.
Figure \ref{f:z} shows the \av \ distributions as a function of redshift. There is only a  gradual decrease in $\tau$ with redshift. This decline is consistent with a scenario where the \av \ distribution is progressively less complete at higher redshift (Malmquist bias for the supernovae suffering from more extinction), very consistent with what previous authors have found.

\subsection{Stellar Mass and Star-formation Rate}

\begin{figure}
\begin{center}
\includegraphics[width=0.5\textwidth]{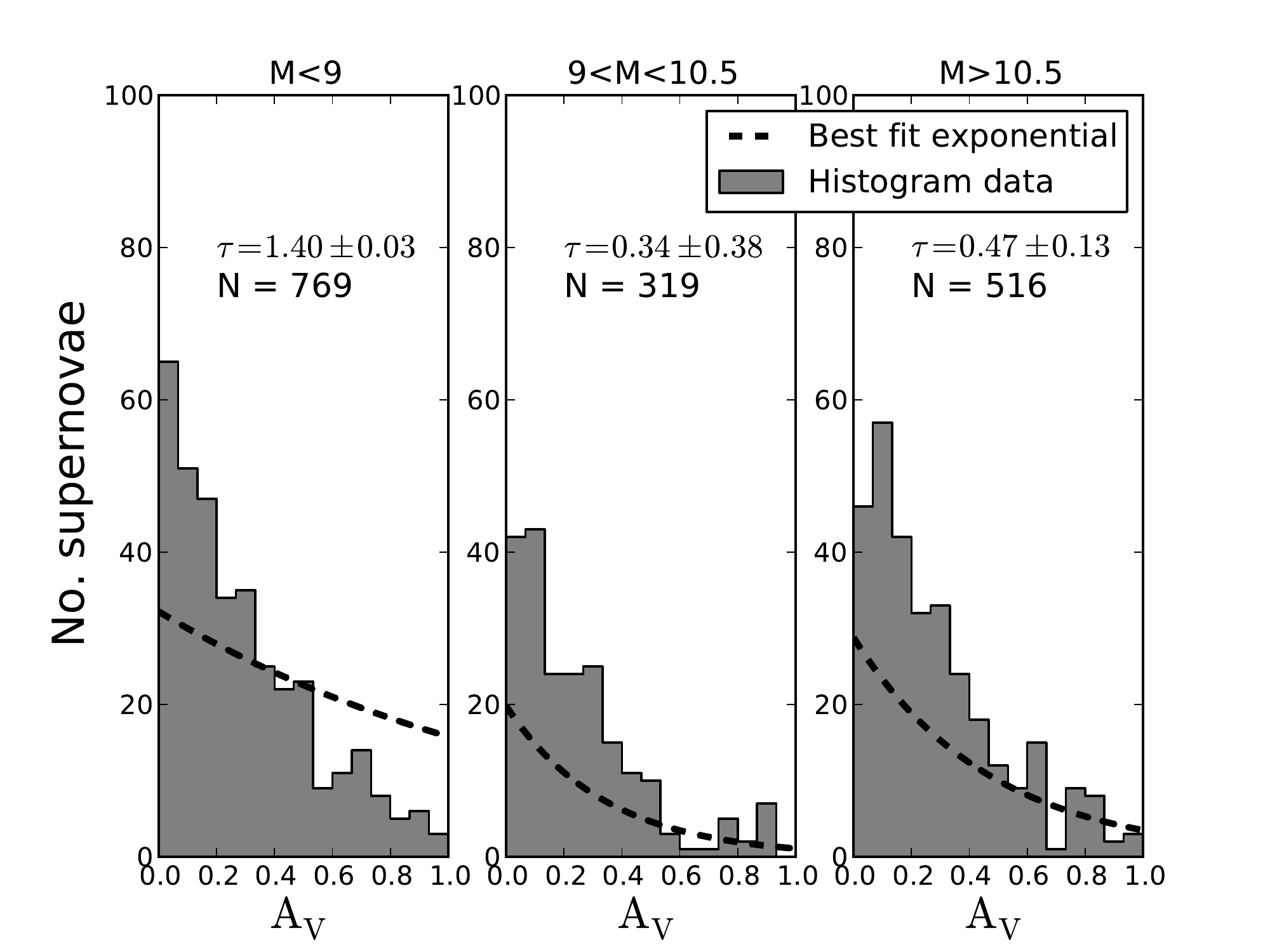}
\caption{The distribution of inclination corrected $A_V$ values for different host galaxy stellar mass (FSPS fit).}
\label{f:mass}
\end{center}
\end{figure}

\begin{figure}
\begin{center}
\includegraphics[width=0.5\textwidth]{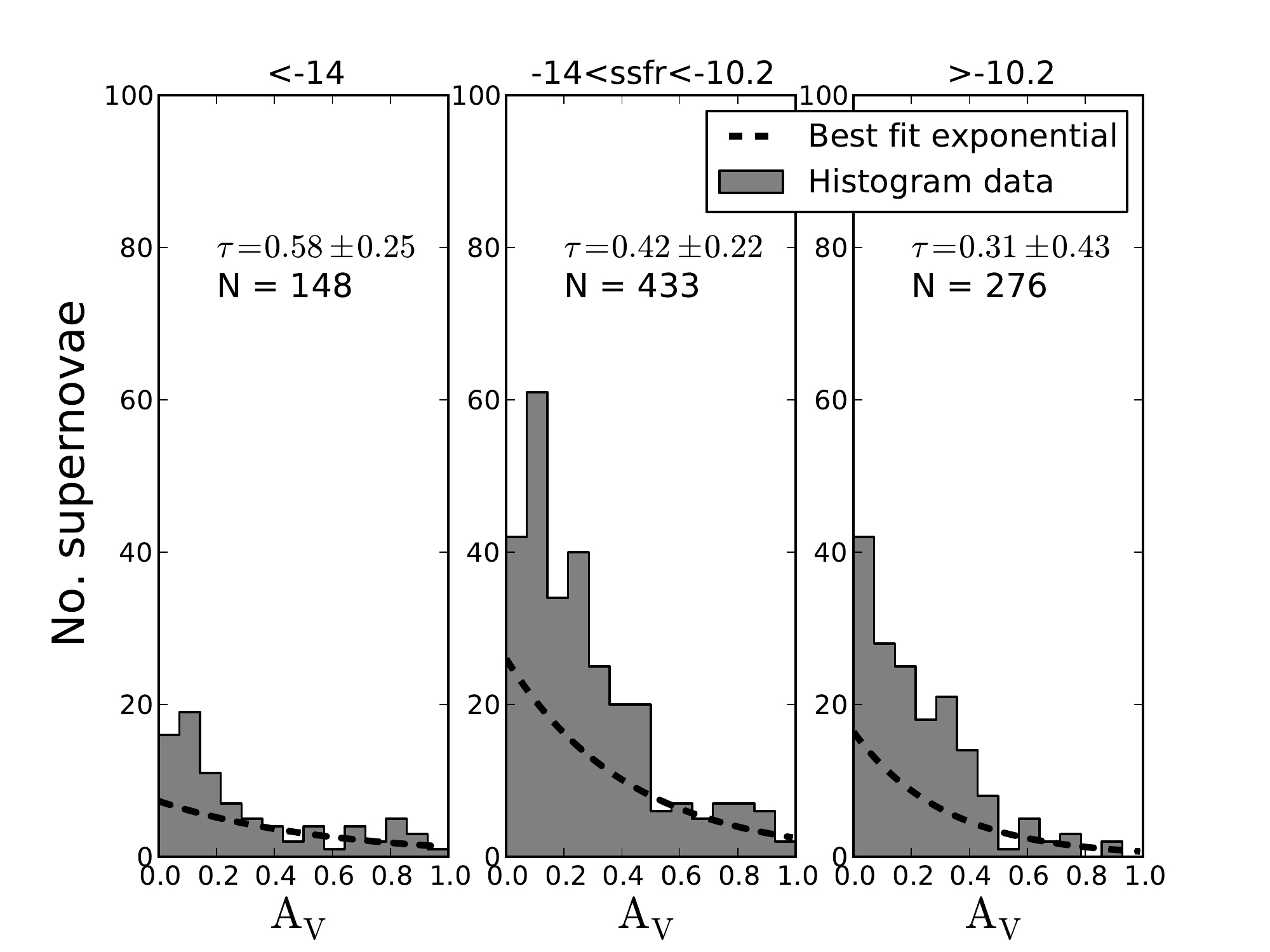}
\caption{The distribution of inclination corrected $A_V$ values for different host galaxy specific star-formation rate (FSPS fit). Bins are in $log_{10}(ssfr)$ in $1/yr$.}
\label{f:ssfr}
\end{center}
\end{figure}

S14 and other authors have already explored the relationship between SN\,Ia color and host galaxy properties such as stellar mass and star-formation rate. The catalog released by S14 includes stellar mass estimates using the {\sc FSPS} \citep{FSPS} and {\sc PEGASE} \citep{Bruzual09} packages and a star-formation estimate from the {\sc FSPS} fit. We use the FSPS fit values.
Figure \ref{f:mass} shows the distribution of $A_V$ values for different stellar masses, and Figure \ref{f:ssfr} the distribution for different specific star-formation levels.
One would expect the distribution to be different as a function of either. 

%
For example, \cite{Dalcanton04} show that the ISM and stellar disks of spiral galaxies are flatter in more massive systems. In smaller galaxies, the ISM (and dusty) disks extends more into the stellar disk, changing the probability that the light from a supernova would encounter significant ISM. Binning the sample into three stellar mass bins however, we note that there is no clear trend (perhaps a slight increase with mass) in the probability a SN\,Ia encountered more or less dust in different mass galaxies in Figure \ref{f:mass}.
This is either because the effect \cite{Dalcanton04} notes happens in much smaller systems or the final $A_V$ distribution does not depend critically on the vertical distribution of the ISM.

Another product of the S14 catalog is the specific star-formation rate of each host galaxy. The distribution of inclination corrected $A_V$ values in Figure \ref{f:ssfr} shows a decline of $\tau$ with increased SSFR.  Most of the SDSS galaxies are around $\rm log_{10}(SSFR) \sim10.2 ~ 1/yr$ which is why the first (quiescent) bin has so few galaxies in it. Contrary to what could be expected from higher star-formation (a wider range of \av \ values due to turbulence in the ISM), the value of $\tau$ {\em declines} with increased star-formation. 
This is contrary to the distribution observed in the interacting galaxy UGC 3995 \citep{Holwerda13b}, where the interaction and accompanying star-formation has removed much of the diffuse component. 
One explanation of any behaviour of decreasing $\tau$ may be that in star-forming galaxies, the dust grains are destroyed or blown out of the line-of-sight by the star-formation prior to supernova ignition \citep[e.g.,][]{Baumgartner13}.

\section{Conclusions}
\label{s:concl}

Using SDSS-II type Ia supernovae as our plumbing lines into the dusty ISM of the host galaxies we can conclude the following: 
\begin{enumerate}
\item The inclination correction we applied to the distribution of extinction values was the simple \av(corrected) = \av(observed) $\rm \times cos(i)$, i.e., $\tau$(corrected) = $\tau$(observed) $\rm \times cos(i)$ for the whole distribution.
The distributions of $A_V$ values become more similar  for the highly and lowly inclined host galaxies after this correction (Figures \ref{f:obs} and \ref{f:cor}). Hence we conclude that this correction is sufficient {\em for the whole distribution}, i.e., 
the differences in $\tau$ between the subsamples after inclination correction are consistent with purely the measurement error. 

 To first order supernovae mostly encounter a thin dusty component of the host galaxy.
\item The distribution of \av \ depends on the radial position within the host galaxy disk. Supernovae in the inner disk exhibit a wider range of \av \ values (Figure \ref{f:Rp}). This change with radius is highly consistent with the picture of disk opacity that has been developed over the last two decades (Figure \ref{f:RAv}).
Therefore the position of the SN in the disk should be considered in future \av \ priors for lightcuve fits. 
\item There is a mild decline in the width of the \av \ distribution with redshift (Figure \ref{f:z}). This is consistent with a selection bias against higher \av \ values at greater distances. 
\item The host galaxy's star-formation level influences the \av \ distribution contrary to our expectations; values of $\tau$ decrease with the level of star-formation (Figure \ref{f:ssfr}). The high star-formation seems to remove higher values of \av \ experienced by SNe. One possibility is that the star-formation clears the ISM prior to the SN ignition. 
\end{enumerate}

We explored the dependence of the \av \ distribution for \sn \  lightcurve fits on host galaxy characteristics. Important factors determining the $A_V$ distribution appear to be the inclination and --to a lesser extent-- the radial position in the disk. 
In future searches for much greater numbers of \sns, specific priors depending on the position in the Host's disk could be of use to decrease the extinction systematic. 
The occulting galaxy program with HST \citep{Holwerda14b} is specifically motivated to provide template $A_V$ distributions of different galaxy masses and types for their use --among others-- as priors for SN\,I a light-curve fits.


%



\section*{Acknowledgements}

We would like to thank the anonymous referee for his or her excellent commentary on the earlier drafts of this manuscript, which helped to improve it.
We acknowledge support from the National Research Foundation of South Africa, specifically the ``Human Capacity Development for Multi-wavelength Astronomy" grant to the Astronomy Department at UCT (P.I. R. Kraan-Korteweg).  The work of B.W. Holwerda is supported by Leiden Observatory.
We thank Cpt. Reynolds for inspiration and useful monologues.
Funding for the SDSS and SDSS-II has been provided by the Alfred P. Sloan Foundation, the Participating Institutions, the National Science Foundation, the U.S. Department of Energy, the National Aeronautics and Space Administration, the Japanese Monbukagakusho, the Max Planck Society, and the Higher Education Funding Council for England. The SDSS Web Site is http://www.sdss.org/.
The SDSS is managed by the Astrophysical Research Consortium for the Participating Institutions. The Participating Institutions are the American Museum of Natural History, Astrophysical Institute Potsdam, University of Basel, University of Cambridge, Case Western Reserve University, University of Chicago, Drexel University, Fermilab, the Institute for Advanced Study, the Japan Participation Group, Johns Hopkins University, the Joint Institute for Nuclear Astrophysics, the Kavli Institute for Particle Astrophysics and Cosmology, the Korean Scientist Group, the Chinese Academy of Sciences (LAMOST), Los Alamos National Laboratory, the Max-Planck-Institute for Astronomy (MPIA), the Max-Planck-Institute for Astrophysics (MPA), New Mexico State University, Ohio State University, University of Pittsburgh, University of Portsmouth, Princeton University, the United States Naval Observatory, and the University of Washington.


\newpage
\appendix

\section{Check with Spectroscopically Confirmed SN\,Ia}

\begin{figure}
\begin{center}
\includegraphics[width=0.5\textwidth]{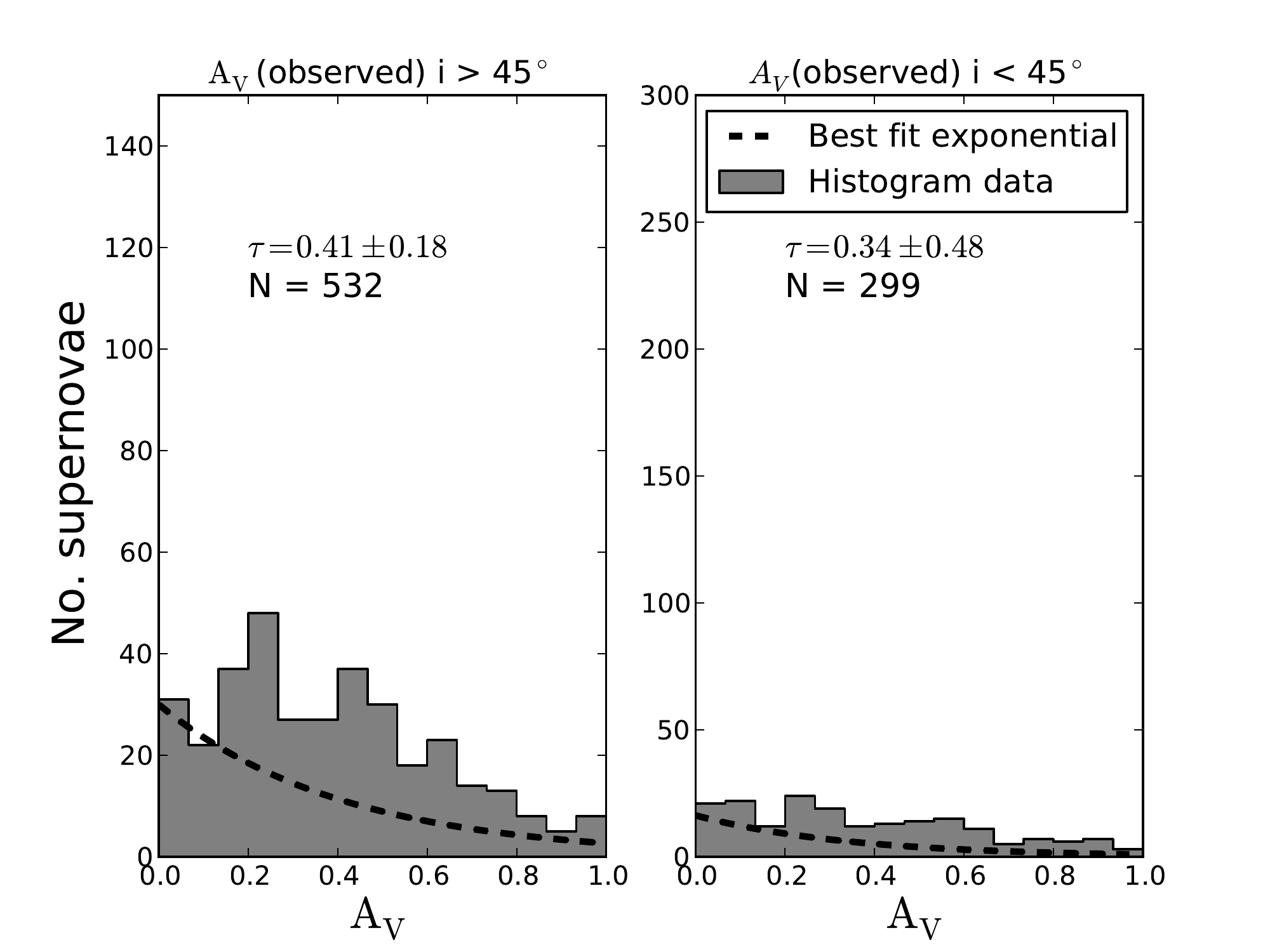}
\caption{{\bf Spectroscopically confirmed SN\,Ia:} The distribution of observed $A_V$ values for the high- ($i>45^\circ$) and low-inclination ($i<45^\circ$) galaxies, before correction for inclination.}
\label{f:conf:obs}
\end{center}
\end{figure}

\begin{figure}
\begin{center}
\includegraphics[width=0.5\textwidth]{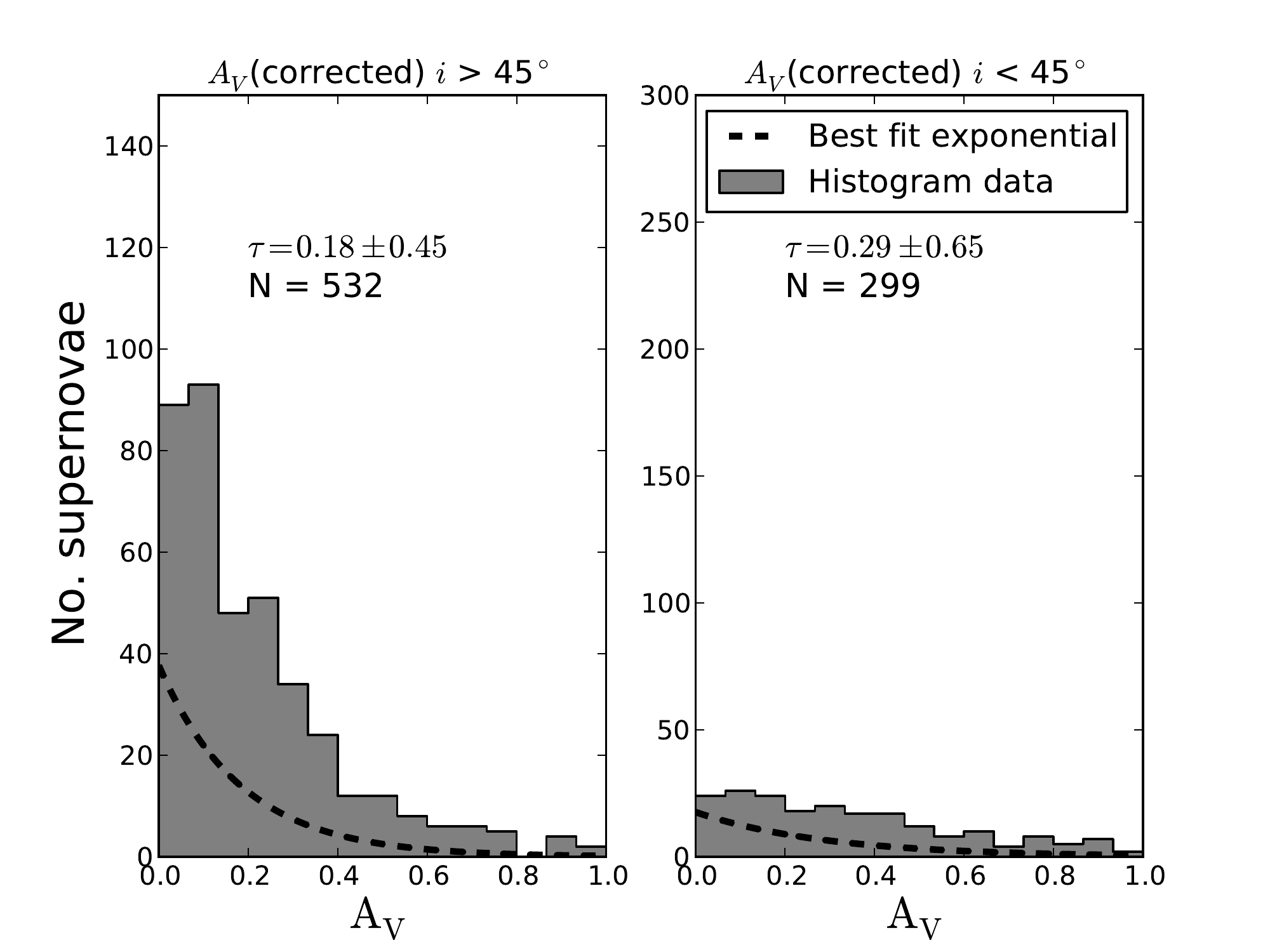}
\caption{{\bf Spectroscopically confirmed SN\,Ia:} The distribution of observed $A_V$ values for the high- ($i>45^\circ$) and low-inclination ($i<45^\circ$) galaxies, before correction for inclination.}
\label{f:conf:cor}
\end{center}
\end{figure}

\begin{figure}
\begin{center}
\includegraphics[width=0.5\textwidth]{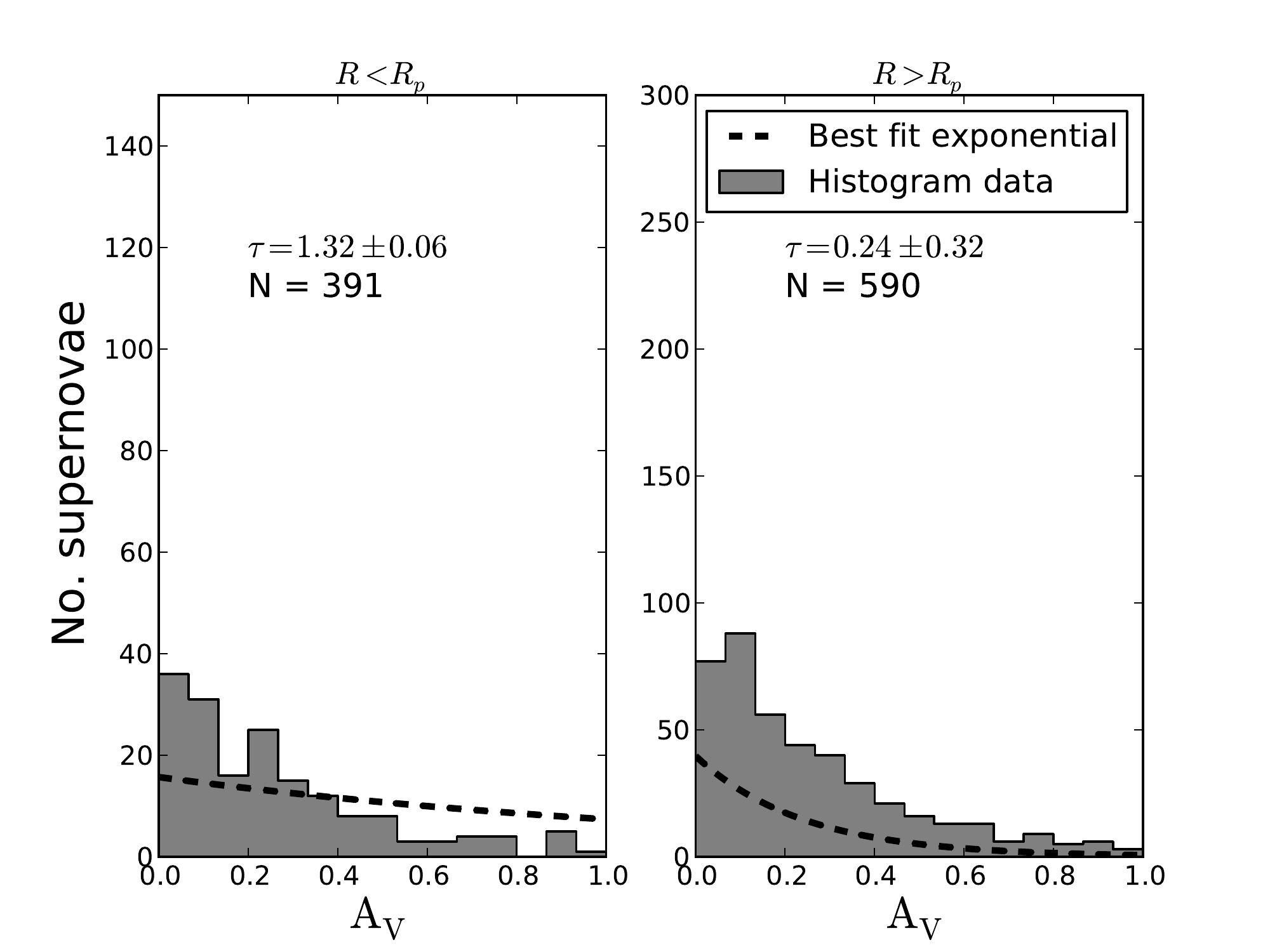}
\caption{The distribution of inclination corrected $A_V$ values for three devisions according to radius, expressed as a fraction of the Petrosian Radius of the host galaxies. Values for $\tau$ decline with radius.}
\label{f:conf:Rp}
\end{center}
\end{figure}

The choice of P(SN\,Ia) $>$ 90\% as the inclusion cut for the SN\,Ia sample is a stringent but ultimately arbitrarily one. We therefore repeat the analysis on those SNe that have been spectroscopically confirmed as SN\,Ia (classifications either 'SNIa', 'pSNIa' or 'zSNIa'). The resulting sample is only 981 objects and thus we can slice it only in courser bins for the comparison.
Figures \ref{f:conf:obs}, \ref{f:conf:cor}, and \ref{f:conf:Rp} are the equivalent figures of Figure \ref{f:obs}, \ref{f:cor}, and \ref{f:Rp} 
for the spectroscopically confirmed sample of SN\,Ia. 


\end{document}